\documentclass[aps,prb,reprint,showpacs,twocolumn,superscriptaddress,floatfix]{revtex4-2}
\usepackage{graphicx}
\usepackage{amssymb}
\usepackage{amsmath}
\usepackage[FIGTOPCAP,nooneline]{subfigure}
\usepackage{overpic}
\usepackage{bbold}

\DeclareMathOperator{\Tr}{Tr}

\begin{document}

\title{Universal Properties of Mesoscopic Fluctuations of the Secondary "Smile" Gap}

\author{J. Reutlinger}

\affiliation{Fachbereich Physik, Universit\"at Konstanz, D-78457 Konstanz, Germany}

\author{L. Glazman}

\affiliation{Department of Physics, Yale University, New Haven CT 06511-8499, USA}

\author{Yu. V. Nazarov}

\affiliation{Kavli Institute of Nanoscience Delft, Delft University of Technology, 2628 CJ Delft, The Netherlands}

\author{W. Belzig}

\affiliation{Fachbereich Physik, Universit\"at Konstanz, D-78457 Konstanz, Germany}



\date{\today}

\begin{abstract}

The energy levels of a quasi-continuous spectrum in mesoscopic systems fluctuate in positions, and the distribution of the fluctuations reveals information about the microscopic nature of the structure under consideration. Here, we investigate mesoscopic fluctuations of the secondary "smile" gap, that appears in the quasiclassical spectrum of a chaotic cavity coupled to one or more superconductors. Utilizing a random matrix model, we compute numerically the energies of Andreev levels and access the distribution of the gap widths. We mostly concentrate on the universal regime $E_{\mathrm{Th}}\gg\Delta$ with $E_{\mathrm{Th}}$ being the Thouless energy of the cavity and $\Delta$ being the superconducting gap. 
We find that the distribution is determined by an intermediate energy scale $\Delta_g$ with the value between the level spacing in the cavity $\delta_s$ and the quasiclassical value of the gap $E_g$. From our numerics we extrapolate the first two cumulants of the gap distribution in the limit of large level and channel number.
We find that the scaled distribution in this regime is the
Tracy-Widom distribution: the same as found by Vavilov at al. [Phys. Rev. Lett. \textbf{86}, 874 (2001)] for the distribution of the minigap edge in the opposite limit $E_{\mathrm{Th}}\ll \Delta$. This leads us to the conclusion that the distribution found is a universal property of chaotic proximity systems at the edge of a continuous spectrum.

\end{abstract}


\maketitle


\section{Introduction}

Normal metals connected to one or more superconductors are subject to the so-called proximity effect \cite{deutscher:69}, which arises due to the penetration of superconducting correlations into the normal metal. Its influence is most striking for the properties of structures with the normal metal part's size being of the order of the superconducting coherence length, which is the length scale for the decay of superconducting correlations inside the normal part. Whereas systems with more than one superconductor involved can host equilibrium supercurrents \cite{josephson:62}, the most eye-catching observable which is strongly modified by the proximity of even a single superconductor is the local density of states (LDOS) in the normal part \cite{belzig:96,levchenko:08}. Besides the emergence of a gap around the Fermi energy $E_F$ \cite{mcmillan:68}, which is known as the \textit{minigap}, another \textit{secondary "smile" gap} was recently reported for a special class of normal structures consisting of a chaotic cavity connected to superconductors via ideally transmitting ballistic channels \cite{reutlinger:14}. 

Such disordered systems are known to exhibit a universal behavior in the sense that the statistical properties of the spectrum in the quasiclassical limit
do not depend on microscopic details of the system, such as the exact distribution of impurities or the exact shape of a ballistic cavity with chaotic scattering at the boundaries. Rather, they are determined by the presence or absence of fundamental symmetries \cite{beenakker:97} in the Hamiltonian of the system. 
This assumption is only true, if the system is sufficiently chaotic. The time an excitation spends inside the normal region before reentering a connector towards a superconductor, the dwell time $\tau_\textrm{dwell}$ must be much larger than the ergodic time $\tau_\textrm{erg}$ required to exploring the whole phase space system\cite{beenakker:97}. The only parameter describing the normal metal properties is thus the energy scale related to the inverse dwell-time: the Thouless energy \cite{Thouless:77},
\begin{align*}
E_{\mathrm{Th}}=\hbar/\tau_\textrm{dwell}.
\end{align*}
The fundamental symmetries in the system are time-reversal symmetry, which can broken by an external magnetic field, and spin rotation symmetry, which is broken in systems where spin-orbit interaction plays a role \cite{Dyson:62,Zirnbauer:11}. In this study, we restrict ourselves to the case where both symmetries are present. 

This universality makes possible a description in terms of random matrices respecting the appropriate symmetries. These matrices are either random Hamiltonians in the description of finite systems, or random scattering matrices in the description of open systems \cite{beenakker:97}. This method - termed Random Matrix Theory (RMT) - turned out to be a powerful tool in the description of average properties \cite{melsen:97}, as well as in the description of mesoscopic fluctuations of average values \cite{vavilov:01}. So far most interest was attributed to the description of the minigap and its statistical properties \cite{beenakker:05,vavilov:03}. In the regime $E_{\mathrm{Th}}\ll \Delta$ the system can be described by an effective Hamiltonian \cite{melsen:96}, whose smallest eigenvalue indicates the gap with an average given by $E_{\textrm{Th}}$. This eigenvalue was found to be distributed according to the universal Tracy-Widom distribution function, which is generally valid for random hermitian matrices \cite{Tracy:94,Tracy:96,vavilov:01} at the edge of a spectral gap.

In this article, we address the opposite regime $E_{\mathrm{Th}}\gg \Delta$. No effective Hamiltonian description of Andreev energy levels below the superconducting gap edge $\Delta$ is possible in this case. The energy positions are found numerically as the roots of a complex determinant equation rather than as the eigenvalues of a matrix. We concentrate on the properties of the secondary "smile" gap in the spectrum, confirm the results of the quasiclassical approach for average density of states, and investigate the averaged mesoscopic fluctuations of the secondary gap width. 

This "smile" gap has been discovered by the authors \cite{reutlinger:14,reutlinger:14_2}  several years ago. We have shown that this gap opens up near the edge of the continuous spectrum $E=\Delta$ in a chaotic cavity coupled to one or more superconductors, in addition to the usual minigap opening at Fermi level. The condition $E_\textrm{Th} \gtrsim \Delta$ is required for this secondary gap.  The secondary gap has a universal behavior at $E_\textrm{Th} \gg \Delta$ \cite{reutlinger:14} where its width is $E_g \approx 0.0147 \Delta^3/E_{\textrm{Th}}^2$. In this limit, the whole subgap density of states has a universal shape not depending on $E_{\textrm{Th}}$. 

We show by a numerical study that this universal behavior holds also for the gap width fluctuations. The distribution is universal and coincides in rescaled units with the universal distribution of the fluctuations of the minigap width \cite{vavilov:01}. Hence, this distribution is characteristic for the superconducting spectral properties close to a gap. 

We stress that this statement is neither straightforward nor obvious. From a mathematical point of view the problem is significantly different, since the energies of the levels cannot be associated with eigenvalues of a random hermitian matrix in the way it was done in~\cite{Tracy:94,Tracy:96} and utilized in~\cite{vavilov:01}. 
There are also differences in the physical setups: the minigap opens at zero energy and is subject to electron-hole symmetry of the spectrum, while the secondary gap is far from zero energy abutting the edge of continuous spectrum at $E = \Delta$, which could affect the level statistics. 

This article is structured as follows: In Section \ref{sec:model} we introduce a random matrix model that describes the setup under consideration and derive the determinant equation for the level energies. In Section~\ref{sec:AvDOS} we evaluate the semi-classical density of states in this model demonstrating the equivalence with the results of Green's function approach implemented in \cite{reutlinger:14}.  In Section~\ref{sec:numbers} we consider the numbers of Andreev levels in energy intervals. Combining numerical and analytical results, we prove that the secondary gap opens at the $N^\mathrm{th}$ Andreev level, $N$ being the number of transport channels opened to the superconductors. This allows us to concentrate on the distribution of energies of this part. In Section \ref{sec:UnivReg} we numerically calculate the distribution of the secondary gap for finite dimensions of the random matrix model and extrapolate to the limit of infinite dimensions to find an accurate correspondence with the universal distribution. 
We conclude in Section~\ref{sec:conclusions}

\section{The model}
\label{sec:model}

In this Section, we motivate and specify the random matrix model in use. 
In general, random matrix models permit evaluation of the average density of states (e.~g.~\cite{melsen:96}) where the results in the limit of the large dimension of the matrices are equivalent to the results of quasiclassical Green's function calculations. Same random matrix models also permit evaluation of mesoscopic fluctuations, for instance, the fluctuations of energy positions of Andreev levels and their statistics \cite{vavilov:01}.

The energy positions of Andreev levels in a 
generic proximitized nanostructure 
 are determined by solutions of Beenakker's determinant equation \cite{beenakker:91,qt}
\begin{equation}
	\label{eq:det1}
	\det\left[1-\hat{S}_N^e(E) \hat{S}^{eh}_A(E)\hat{S}_N^h(E)\hat{S}^{he}_A(E)\right]=0.
\end{equation}
They are thus determined by an energy-dependent electron scattering matrix $\hat{S}_N^e(E)$ inside the normal region (N). This is an $N\times N$ matrix in the space of all transport channels coming into or going out of the nanostructure.
The scattering matrix for holes  is related to that of electrons, $\hat{S}_N^h(E)=\hat{S}_N^{e*}(-E)$. The transport channels are opened to superconducting terminals where electrons are converted into holes and vice versa. This is described by Andreev scattering matrices $\hat{S}_A^{eh,he}(E)$ that can be chosen to be diagonal, 
$(\hat{S}_A^{eh,he}(E))_{ii} =\exp[-i \arccos(E/\Delta_i)] \exp(\pm i \phi_i)$, $\Delta_i,\phi_i$ being the modulus and phase of the superconducting order parameter in a terminal the channel $i$ belongs to. For the same phase and modulus in all terminals, 
$(\hat{S}_A^{eh,he}(E)) =\exp[-i \arccos(E/\Delta)]$, the Andreev scattering matrices can be just replaced by an energy-dependent phase factor.

If the nanostructure is sufficiently short so that $E_{\rm Th} \gg \Delta$, one can neglect the energy dependence of scattering matrix. For ballistic transport in a chaotic cavity, $\hat{S}_N$ can be taken as a random member of one of the circular ensembles of RMT \cite{blumel:90,beenakker:97}. In this work, we imply time reversibility and, hence, assume a time-reversible scattering matrix that is a member of the circular orthogonal ensemble. However, the existence of smile gaps implies $E_{\textrm{Th}} \gtrsim \Delta$, so the energy dependence of the scattering matrix cannot be neglected. To model the situation, 
we adopt a Hamiltonian representation of the scattering matrix proposed in \cite{Weidenmueller} and utilized in \cite{frahm:96,beenakker:05} in superconducting context,
\begin{equation}
\hat{S}_N = 1 - 2\pi i \hat{W}^\dagger \left(E -\hat{H} + i \pi \hat{W}\hat{W}^\dagger\right)^{-1} \hat{W} 
\end{equation}
Here, the Hamiltonian $\hat{H}$ is an $M\times M$ Hermitian matrix describing the electron states in an isolated cavity. For  a chaotic cavity, this Hamiltonian is a member of the Gaussian orthogonal ensemble, whose probability distribution is defined by \cite{mehta:91}
\begin{equation}
	\label{eq:gaussian}
	\mathcal{P}(\hat{H})\sim \exp(-\frac{\pi^2}{4\delta_s^2 M} \Tr{\hat{H}^2}).
\end{equation}
$\delta_s$ being the mean level spacing of the isolated cavity.

The $N \times M$ matrix $\hat{W}$ describes the connection between the electron states in the cavity and terminals via $N$ transport channels with transmissions $T_n$. It is defined upon unitary transformations in the spaces of channels and states. The transmission coefficients of $N$ transport channels are related to N eigenvalues 
of $\hat{W}^\dagger \hat{W}$,
\begin{equation}
	\label{eq:wthruT}
w_n = \frac{M \delta_s}{\pi^2 T_n}  (2-T_n - 2\sqrt{1-T_n})
\end{equation}
For ballistic connectors, $T_n=1$, $w_n = M\delta_s/\pi^2$.  The simplest way to choose $\hat{W}$ is to set $W_{nm} = \delta_{nm} \sqrt{w_n}$.

It has been shown in \cite{frahm:96,beenakker:05} that Eq.~(\ref{eq:det1}) with this scattering matrix can be transformed to
\begin{equation}
	\label{eq:det2}
	\det({E}\hat{1}-\hat{\mathcal{H}}+\hat{\mathcal{W}}(E))=0.
\end{equation}
Here, an extra $2\times2$ Nambu structure has been introduced and 
the matrices $\hat{\mathcal{H}}$ and $\hat{\mathcal{W}}$ are defined as
\begin{gather*}
	\hat{\mathcal{H}}=H \hat{\sigma}_3, \\ 
	\hat{\mathcal{W}}(E)=\frac{\pi}{\sqrt{\Delta^2- E^2}} \begin{pmatrix} E W W^{\dagger} & \Delta W W^{\dagger}\\ \Delta W W^{\dagger} & E W W^{\dagger} \end{pmatrix}.
\end{gather*}
 While Eq. (\ref{eq:det2}) may resemble an eigenvalue equation defining a spectrum of a Hamiltonian, it is not one since the Hamiltonian $\hat{\mathcal{H}}$ of the isolated normal part is accompanied by the energy-dependent selfenergy $\hat{\mathcal{W}}(E)$. This significantly complicates the numerical solution. 

For this model, the $2M \times 2M$ Green's function can be defined as follows:
\begin{equation}
\label{eq:GF}
\hat{\mathcal{G}}(z) = \frac{\hat{\mathbb{1}}}{z\hat{\mathbb{1}}-\hat{\mathcal{H}}+\hat{\mathcal{W}}(z)}
\end{equation}
 After the averaging over the random $\hat{\mathcal{H}}$ (\ref{eq:gaussian}), it is diagonal in the space of electron states. 
 
The average density of states can be computed from this Green's function as 
\begin{align}
	\label{eq:DOS}
	\rho(E)=& \sum_n \langle \delta(E -E_n )\rangle \nonumber \\
	& =-\frac{1}{\pi} \Im \langle \Tr \left[(\hat{\mathbb{1}}+d\hat{\mathcal{W}}/dE) \hat{\mathcal{G}}(E+i\delta^{+})\right]\rangle,
\end{align}
the first sum is a sum over Andreev levels.
The factor $(\hat{\mathbb{1}}+d\hat{\mathcal{W}}/dE)$ has to be incorporated  to account for evanescent propagation of Andreev states into the superconducting terminals. We can also define the local DOS in the normal region that can be immediately measured by a tunnel contact connected to the normal part. In this case, each Andreev bound state is weighted with probability $P_n$ to be in the normal region. The local DOS is expressed by similar relation without the factor,
\begin{align}
	\label{eq:DOSloc}
	\rho_{\rm loc}(E)=& \sum_n \langle P_n\delta(E -E_n )\rangle \nonumber \\
	& =-\frac{1}{\pi} \Im \langle \Tr \left[\hat{\mathcal{G}}(E+i\delta^{+})\right]\rangle,
\end{align}

In our numerics, the energies of Andreev levels are computed by finding the roots of the determinant given Eq. (\ref{eq:det2}). For Andreev levels close to the secondary gap\cite{reutlinger:14} with energies $E_\textrm{A} \lesssim \Delta$, there is no obvious possibility to reduce this problem to an eigenvalue problem of an effective Hamiltonian. This was possible in similar studies \cite{vavilov:01} of the level statistics of the minigap in the limit $E_{\textrm{Th}} \ll \Delta$ and appeared to simplify the calculations greatly. 

For the model under consideration, 
$E_{\rm Th} = N \delta_s/(2\pi)$. To account for constant density of normal electron states at $E <\Delta$ the width of the spectrum of $\hat{H}$, $\simeq M \delta_s$, should significantly exceed $\Delta$. This is why the semiclassical regime with $E_{\rm Th} \gg \Delta$ implies $M \gg N \gg 1$.

\section{Average density of states}
\label{sec:AvDOS}
In this Section, we evaluate the average DOS for the RMT model formulated in the previous section. We will  show explicitly the equivalence of the results with those obtained by the semiclassical Green's function method employed in\cite{reutlinger:14}. We thus prove the occurrence of the secondary gap in an RMT model.

 In our treatment of the RMT model, we follow the approach by Melsen et al. \cite{melsen:96} modifying it for a calculation that is valid in the whole energy interval $\lbrack 0, \Delta \rbrack$ and gives both the full and local densities of states in the system. We will work in  the limit $M\gg N \gg 1$ using the perturbation expansion in $1/M$  to average over the Hamiltonians of the Gaussian orthogonal ensemble. We introduce the average Green's function that is a matrix in Nambu space, and upon the sign change of the elements, it is equivalent to the quasiclassical Green's function used in \cite{reutlinger:14}:
\begin{displaymath}
	\hat{G}(z)=\frac{\delta_s}{\pi} \begin{pmatrix} \Tr \mathcal{G}_{11} & \Tr \mathcal{G}_{12} \\ \Tr \mathcal{G}_{21} & \Tr \mathcal{G}_{22} \end{pmatrix}
\end{displaymath}
$ \mathcal{G}_{ij}$ are the $M \times M$ subblocks of the matrix $\mathcal{G}$ from Eq.~(\ref{eq:GF}). 

To compute this matrix, we employ a traditional self-consistent Born approximation valid for $M \gg 1$,
\begin{align}
\label{eq:GF_eq_1}
\hat{G}(z) &=\frac{\delta_s}{\pi} \sum_{n=1}^{M} \frac{ 1}{(\hat{G}^{(0)}_n)^{-1} - \hat{\Sigma}}, \\
&  (\hat{G}^{(0)}_n)^{-1} = z + \frac{\pi w_n}{\sqrt{\Delta^2-z^2}} \begin{pmatrix}z & \Delta \\ \Delta & z \end{pmatrix}; \\
& \Sigma = \frac{M\delta_s}{\pi} \begin{pmatrix} G_{11} & -G_{12} \\ -G_{21} & G_{22} \end{pmatrix}
\end{align}

We note the symmetry of the elements\cite{melsen:96}:
\begin{displaymath}
	G_{11}=G_{22},~
	G_{12}=G_{21},~
	G_{12}^2=1+G_{11}^2
\end{displaymath}
To proceed, we implement a simple model of a transmission distribution where all channels have the same transmission $T$ and therefore the same $w$ given by Eq. \ref{eq:wthruT}. With this, the sum over $n$ can be readily computed and 
 Eq. \ref{eq:GF_eq_1} can be reduced to 
\begin{equation}
	\label{eq:GF_eq_2}
	z G_{12}\frac{\sqrt{1-z^2}(T-2)-T(zG_{11}+G_{12})}{G_{11}+z G_{12}}=\frac{TN\delta_s}{2\pi}.
\end{equation}
This model has been also used in Ref.~\onlinecite{reutlinger:14}.
In the following steps, we introduce the Thouless energy $E_{\textrm{Th}}=TN\delta_s/(2\pi)$, express $G_{12}$ via $G_{11}$ and identify $G_{11}=-ig$. With this, Eq.~(\ref{eq:GF_eq_2}) becomes precisely equal to Eq.~(2) in Ref.~\onlinecite{reutlinger:14}. We have therefore demonstrated the equivalence of RMT and quasiclassical Green's functions approaches.

With this, we can compute the local DOS. To evaluate the full DOS, we  have to account for the term arising from the coupling to the superconductors $\sim d\hat{\mathcal{W}}/dE$, that is expressed as
\begin{displaymath}
\Tr\left[(d\hat{\mathcal{W}}/dE) \hat {\mathcal{G}}\right] \propto \Tr_N [\mathcal{G}_{11}]+(E/\Delta) \Tr_N [\mathcal{G}_{12}] .
\end{displaymath}
Here $\Tr_N(...)$ indicates a trace over the first N diagonal components in the particular block $\mathcal{G}_{ij}$, where the $w_n$ are non-zero. We define a Green's function that involves only this summation:
\begin{displaymath}
\hat{g}=\frac{\delta_s}{\pi} \begin{pmatrix} \Tr_N \mathcal{G}_{11} & \Tr_N \mathcal{G}_{12} \\ \Tr_N \mathcal{G}_{21} & \Tr_N \mathcal{G}_{22} \end{pmatrix}
\end{displaymath}
which is readily expressed as 
\begin{align}
	\label{eq:GF_eq_3}
	\hat{g}(z) &=\frac{\delta_s}{\pi} \sum_{n=1}^{N} \frac{ 1}{(\hat{G}^{(0)}_n)^{-1} - \hat{\Sigma}}. 
\end{align}
In the model of a constant $T$, we find 
\begin{displaymath}
	g_{11}(z)+\frac{z}{\Delta} g_{12}(z)=-\frac{1}{w \pi} \frac{\Delta z}{\sqrt{\Delta^2-z^2}} G_{12}(z),
\end{displaymath}
$w$ being given by Eq.~\ref{eq:wthruT}. 
Substituting this into Eq. (\ref{eq:DOS}) we finally find for the full DOS:
\begin{equation}
	\label{eq:DOS2}
	\rho(E)=-\frac{2}{\delta_s} \Im \left[G_{11}(E)-\frac{E\Delta}{\Delta^2-E^2}G_{12}(E)\right].
\end{equation}
In addition to the contribution to the full DOS from the part of the Andreev states located inside the normal part, that is proportional to $G_{11}$, there is also a contribution from the parts  leaking into the superconductors. This contribution is related to the anomalous component of the Green's function $G_{12}$, which is caused by the proximity of the superconductors. This contribution is energy-dependent and diverges for energies close to $\Delta$, this reflects the fact, that Andreev states with energies $E_{\textrm{A}}\approx \Delta$ are mainly located in the superconductors.

\section {Number of Andreev levels in Energy Intervals}
\label{sec:numbers}
\begin{figure}[t] 
\vspace{10mm}
\begin{overpic}[width=0.95\columnwidth,angle=0]{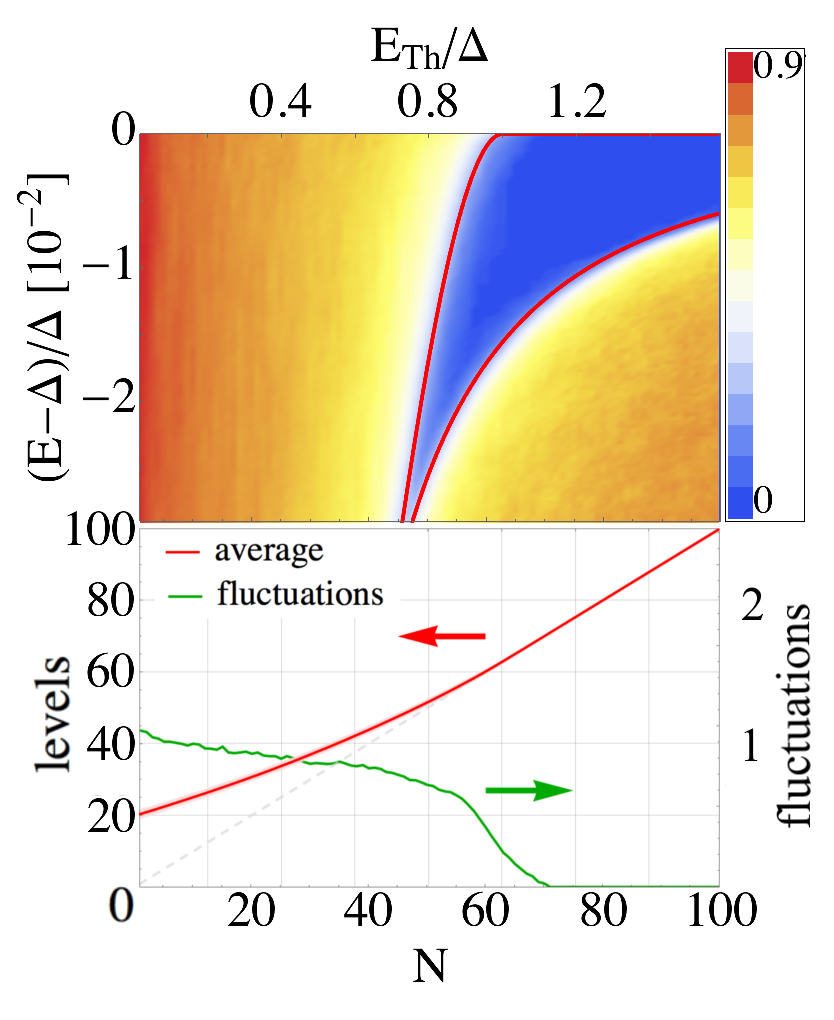}
\put(3,95){\makebox(0,3){$(a)$}}
\put(3,48){\makebox(0,3){$(b)$}}
\end{overpic}
\caption{\label{fig:1} (a) 
The fluctuations of the number of Andreev levels in an energy interval $\lbrack 0, E \rbrack$ as a function of number of channels $N$. The solid red lines indicate the secondary gap edges in quasiclassical approximation.The fluctuations are strongly suppressed but non-zero in the quasiclassically forbidden domain. (b) The fluctuations and average number of Andreev levels in the interval $\lbrack 0, \Delta \rbrack$. For $E_{\textrm{Th}} > \Delta$ the fluctuations are suppressed and the number of levels is pinned to $N$. For $N \to 0$, the cavity is uncoupled from the superconductors and the average number of levels is $2 \Delta/\delta_s$.}
\end{figure}

Before addressing the fluctuations of the gap, we need to know how many Andreev levels are situated in the energy interval between the minigap and the secondary gap so we know which levels are separated by the secondary gap. In this Section, we present our numerical and analytical results that concern the number of Andreev levels in energy intervals.


In our numerical calculations, we fix the number of levels inside the cavity to $M=500$ and the level spacing to $\delta_s=0.1\Delta$.  We generate 2000 pseudo-random Hamiltonian matrices with the distribution given by Eq. (\ref{eq:gaussian}). For each realization, we solve Eq. (\ref{eq:det2}) to find the energy levels at various $E_{\textrm{Th}}$ that is tuned by varying the number of channels $N$ opened to the superconductor. 
Changing $N$ from 0 to 100 corresponds to $E_{\textrm{Th}}$ being increased from 0 to approximately $1.6 \Delta$. 

The results are presented in Fig. \ref{fig:1}. The color plot in Fig. \ref{fig:1}  shows the fluctuations of the number of levels in an interval $\lbrack 0, E \rbrack$ as a function of  $E_{\textrm{Th}}$ (x-axis) and as a function of energy $ E $ (y-axis) close to the gap edge. Generally, the fluctuations are of the order of $1$ as expected from the spectral rigidity of the random matrices. The solid red lines show the boundaries of the secondary gap found from the above quasiclassical calculations. 
We see the strong suppression of the fluctuations in the quasiclassically forbidden domain yet they are still non-zero 
corresponding to single isolated Andreev levels that enter the domain as fluctuations. These fluctuations decrease at larger Thouless energies $E_{\textrm{Th}} > \Delta$. 
 
Fig. \ref{fig:1} (b) shows detailed results for the energy interval $\lbrack 0, \Delta \rbrack$. Here, we plot the fluctuations of the number of Andreev levels along the average number of levels. 
Till  $E_{\textrm{Th}}\approx 0.8 \Delta$ the number of Andreev levels fluctuates at usual scale. This indicates that in this regime the discrete spectrum of Andreev levels is not separated from the continuum, a level can merge with the continuum or come back as a result of a fluctuation. At bigger  $E_{\textrm{Th}}$ the fluctuations decrease rapidly. For the parameters $M$, $\delta_s$ chosen we see no fluctuations above  $E_{\textrm{Th}} = 1.2 \Delta$ and the number of the bound states is exactly $N$ corresponding to the number of open channels. This indicates a perfect separation of continuous and discrete spectrum. These values of $E_{\textrm{Th}}$ correlate with  $E_{\textrm{Th}} = \Delta$ at which the upper boundary of the secondary gap merges with the continuum. The same tendency we see in the $E_{\textrm{Th}}$ dependence of average number of levels. At $E_{\textrm{Th}}=0$, where the normal region is isolated from the superconductors, the average number of levels is given by $2 \Delta/\delta_s$ (The factor $2$ arises since both electron and hole states are counted as Andreev levels). Upon increasing $N$, the average number grows slower than $N$, becomes equal to $N$ at about $E_{\textrm{Th}} = \Delta$ and does not change any further being pinned to the number of channels.
 
This suggests that the secondary gap opens up exactly between the $N^\mathrm{th}$ and $N+1^\mathrm{st}$ level.

\begin{figure}[t] 
 \includegraphics[width=0.95\columnwidth,angle=0]{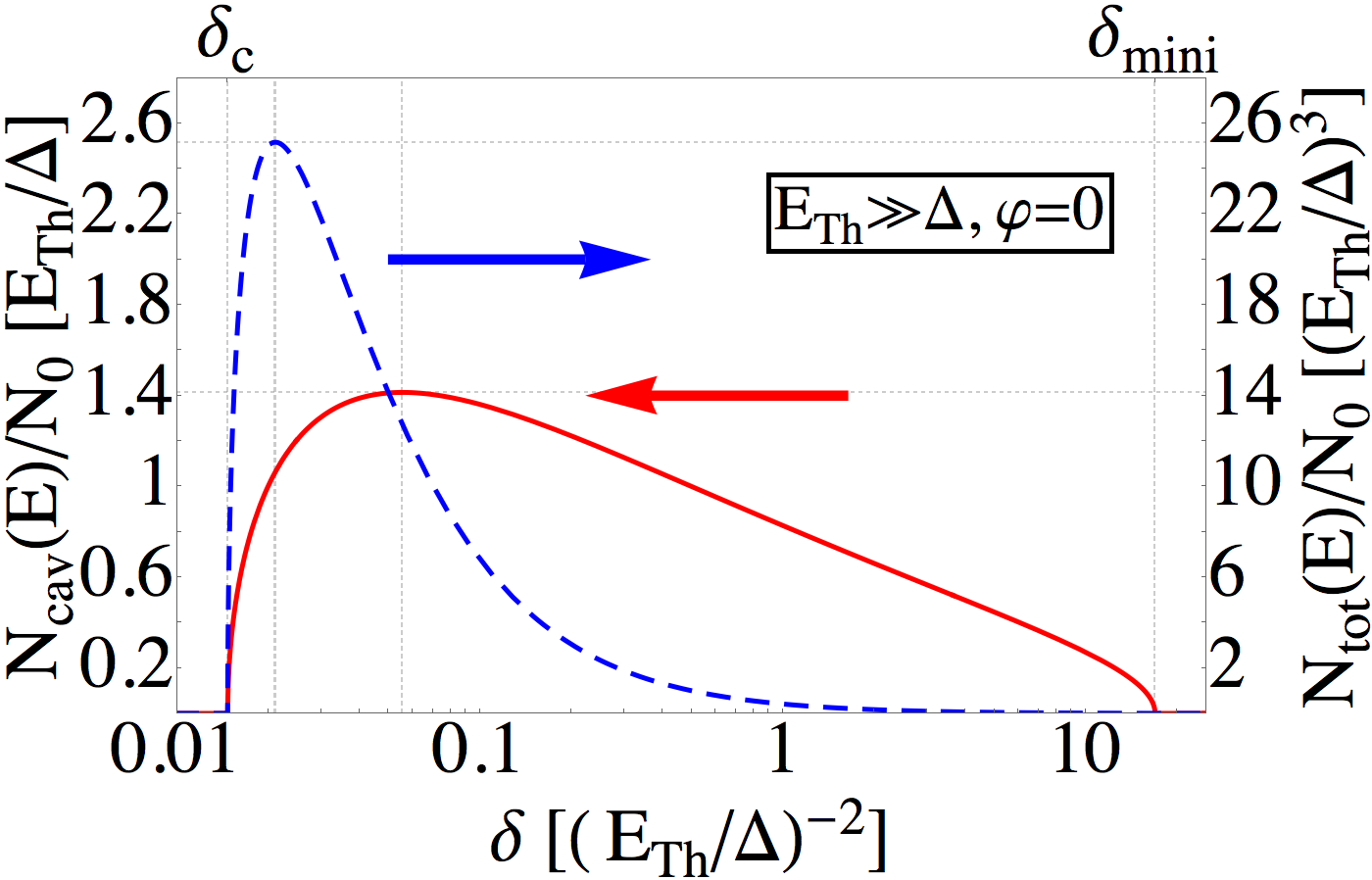}
\caption{\label{fig:2} Comparison of the local DOS inside the normal cavity (solid red line) and the full density of states, which accounts for Andreev levels localized inside the superconductors (dashed blue line) in the limit $E_{\textrm{Th}}/\Delta \to \infty$ and for ballistic coupling ($T=1$). The difference is most important at Andreev energies $E_\textrm{A}$ close to $\Delta$.}
\end{figure}

This result can also be obtained from the quasiclassical density of states. We consider here the limit $E_{\textrm{Th}} \gg \Delta$. It was shown in \cite{reutlinger:14} that in this limit all quantities in Eq. (\ref{eq:GF_eq_2}) can be rescaled with factors $(E_{\textrm{Th}}/\Delta)^k$ of appropriate power $k$ in a way that the Thouless energy drops out of this equation and the rescaled density of states has a universal shape. In order to calculate the total number of Andreev levels from this result one has to pay special attention to the fact that in this limit all Andreev levels have energies close to $\Delta$, given by $\Delta-E_\textrm{A}\sim \Delta^3/E_\textrm{Th}^2$. These states are mostly localized in the superconductors. In terms of Eq.~(\ref{eq:DOS2}) for the full density of states this means that $E\Delta/(\Delta^2-E^2) \gg 1$ is the dominant contribution. To find the total number of Andreev levels, the anomalous Green's function $G_{12}$ is expressed via $G_{11}$ via the normalization condition, then this expression is expanded in $1/G_{11}\sim 1/E_{\textrm{Th}}$ which is small in the limit $E_{\textrm{Th}} \to \infty$:
\begin{displaymath}
	G_{12}=-\sqrt{1+G_{11}^2}\approx-G_{11}-1/(2G_{11})
\end{displaymath}
Introducing rescaled quantities $(\Delta-E)/\Delta=x(\Delta/E_{\textrm{Th}})^2$ and $G_{11}=y E_{\textrm{Th}}/\Delta$, the leading order contribution to the full DOS becomes
\begin{equation}
	\label{eq:DOS3}
	\rho(E)=-\frac{N T}{\pi E_{\textrm{Th}}} \left ( \frac{E_{\textrm{Th}}}{\Delta}  \right )^3 \Im \left (\frac{y}{2x}\right ).
\end{equation}
The difference between the full DOS and the local DOS  is shown in Fig.~\ref{fig:2}. The scaling of both curves with $E_{\textrm{Th}}/\Delta$ is different and for large $E_{\textrm{Th}}$ the local DOS can be neglected in comparison with the full one. We make use of the solution for $y$, found in \cite{reutlinger:14} and \cite{reutlinger:14_2} and integrate Eq.~(\ref{eq:DOS3}) from the minigap edge $\delta_{mini}$ to the secondary gap edge $\delta_c$. In the universal limit $E_{\textrm{Th}}/\Delta \to \infty$ these boundaries are given by
\begin{align*}
	\delta^b_{mini}=\left( 17/2+6\sqrt{2}\right)(\Delta/E_{\textrm{Th}})^2,\\
	\delta^b_{c}=\left( 17/2-6\sqrt{2}\right)(\Delta/E_{\textrm{Th}})^2.
\end{align*}

For a more general case of constant contact transmission $T$, these boundaries are computed in\cite{reutlinger:14_2}. For tunnel contacts, they are given by
\begin{gather*}
	\delta^t_{mini}=8(\Delta/E_{\textrm{Th}})^2,\\
	\delta^t_{c}=0,
\end{gather*}
there is no secondary gap for the tunnel case.
 Both for ballistic and  tunnel case, the integration can be done analytically.  For the case of general transmission, the integration has to be done numerically. In any case, the integration yields  exactly $N$ Andreev levels for any value of  $T$. This perfectly agrees with our numerical calculations. 

The statistics of the secondary gap is thus the statistics of the level spacing between the $N^\mathrm{th}$ and $N+1^\mathrm{st}$  Andreev level.

\section{Statistics of the secondary gap}
\label{sec:UnivReg}

In this Section, we present the results of numerical investigation of the distribution of the secondary gap. We concentrate on the limit of large Thouless energies, where the average density of states is given by an analytical expression and has a universal shape. In this limit, the upper gap edge is fixed to the edge of the continuum spectrum $\Delta$ and, as shown in the previous section, no levels from the continuum enter the gap region. Thus the gap is given by the energy of the highest Andreev level. For $M \gg N \gg 1$ we expect a universal behavior not only for the average DOS, but also for the fluctuations of the gap. In previous complementary studies \cite{vavilov:01} of the minigap statistics in the limit of small $E_{\textrm{Th}}$, the latter condition  was automatically satisfied by reducing the problem to an eigenvalue problem of an effective Hamiltonian. In this case the only two parameters affecting the result are $M$ and $N$. For the case in hand, the situation is slightly more complicated, since the condition $E_{\textrm{Th}}/\Delta \gg 1$ is not  fulfilled automatically. There are three parameters to vary: the number of levels $M$, the number of channels $N$ and furthermore  the level spacing inside the normal part $\delta_s$. They have to be chosen such that the condition $E_{\textrm{Th}}=N\delta_s/(2\pi)\gg \Delta$ is fulfilled. 

Like in case of \cite{vavilov:01}, the energy scale governing the fluctuations should be the same as the one for the average density of states. In the limit $M \gg N \gg 1$ where the quasiclassical calculation is valid we expand the universal result for the full DOS below the gap in a series to find in lowest order a square-root behavior near the gap egde, 
\begin{displaymath}
	\rho(E)\approx \frac{1}{\pi} \sqrt{\frac{E_c-E}{\Delta_g^3}},
\end{displaymath}
where the energy scale $\Delta_g$ is given by
\begin{displaymath}
 \Delta_g=c E_\textrm{sm}^{4/3} \delta_s^{2/3}/\Delta.
\end{displaymath} 
Here, $E_{sm}=\Delta-E_c$ is the secondary smile gap and $c\approx 1.19$ is a numerical prefactor. 
This energy scale is thus of the order of the energy spacing between the last Andreev levels at the gap edge. 
The definition of $\Delta_g$ is similar to the definition in \cite{vavilov:01}, with the minigap energy $E_\textrm{mini}$ being replaced by $E_\textrm{sm}^4/\Delta^3$. There are two reasons for this difference. The first reason is as follows: for the minigap in the limit $E_{\textrm{Th}}/\Delta \ll 1$ Andreev levels are close to 0 and the contribution from the superconductors to the DOS is negligible. It is thus sufficient to consider the local DOS in the normal part. For the secondary gap it is essential to consider the full density of states to determine $\Delta_g$, since the secondary gap is situated close to $\Delta$ and Andreev levels are mostly localized in the superconductors. The second reason is the different scaling of $E_\textrm{sm}$ with $E_\textrm{Th}$ for $E_\textrm{Th} \gg \Delta$ .

Using the relations $E_{sm}\sim \Delta^3/E_\textrm{Th}^2$ which is valid for $E_\textrm{Th}\gg \Delta$ and $E_\textrm{Th}\sim N\delta_s$, we find that $E_{sm}/\Delta_g \approx N^{2/3}$ in this regime. Thus in universal units of $\Delta_g$ the smile gap $E_\textrm{sm}$ only depends on $N$. In the following consideration, the energies are normalized to $\Delta_g$ and expressed as to $x=(\Delta-E)/\Delta_g$. The value for $\Delta_g$ is universal only in the limit $M \gg N \gg 1$. In the following, we use this definition for finite values of $M$ and $N$. It turns out that this mostly affects the average position of the gap edge, while the distribution only weakly depends on the exact values of $M$ and $N$ .

\subsection{Gap distribution for $M/N = 5$}

\begin{figure}[t]
\begin{overpic}[width=0.98\columnwidth,angle=0]{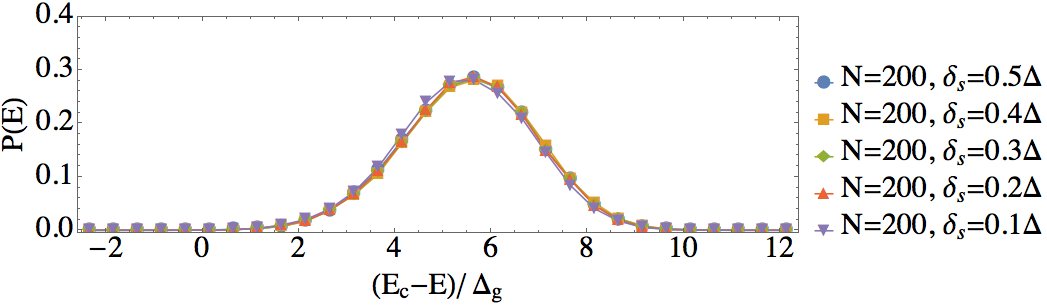}
\put(3,33){\makebox(0,3){$(a)$}}
\end{overpic}
\begin{overpic}[width=0.98\columnwidth,angle=0]{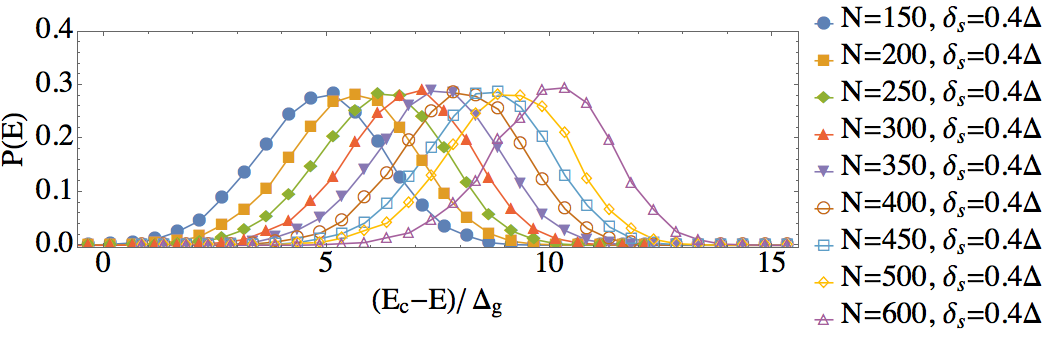}
\put(3,33){\makebox(0,3){$(b)$}}
\vspace{0.2cm}
\end{overpic}
\begin{overpic}[width=0.95\columnwidth,angle=0]{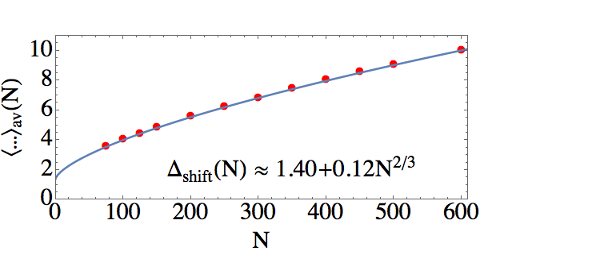}
\put(1,43){\makebox(0,3){$(c)$}}
\end{overpic}
\begin{overpic}[width=0.98\columnwidth,angle=0]{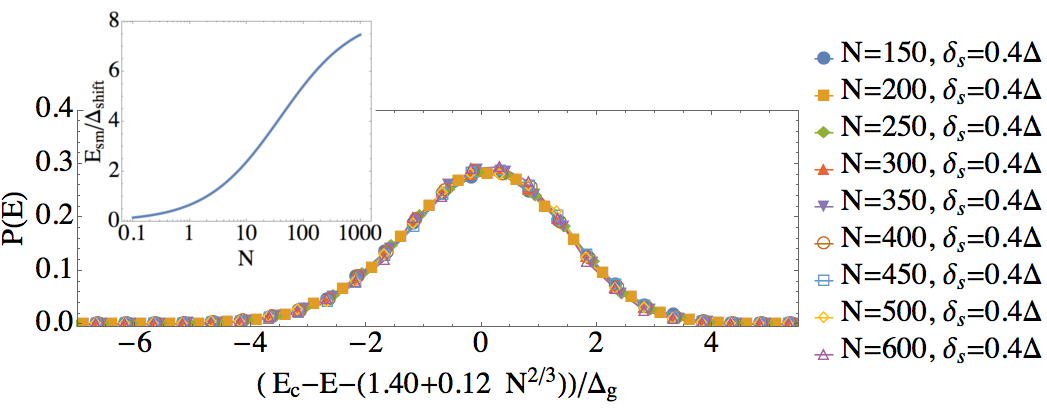}
\put(3,39){\makebox(0,3){$(d)$}}
\end{overpic}
\caption{\label{fig:3} The distributions of the energy of the highest (N-th) Andreev level at a fixed ratio $M/N=5$. (a) The number of channels is fixed to $N=200$ while the level spacing $\delta_s$ is changing from $\delta_s=0.1 \Delta$ to $\delta_s=0.5 \Delta$. The distributions are identical but shifted by $\approx 5 \Delta_g$ with respect to the quasiclassical edge. (b) The level spacing is fixed to $\delta_s=0.4\Delta$. The distributions are computed for different values of $N$ ranging from $N=150$ to $N=600$. (c) The fit of $N$-dependence of the average values of these distributions. (d) Shifting the distributions with the values obtained from the fitting model confirms the agreement of the distributions for different $N$. The inset in this plot shows the ratio of the quasiclassical gap and the shift at $M/N=5$.}
\end{figure}

Here we consider the gap distribution for a fixed finite ratio $M/N=5$. We take $N=200$ and five different values for $\delta_s$ ranging from $\delta_s=0.1 \Delta$ to $\delta_s=0.5 \Delta$. We calculate the distribution of the gap  from $10^4$ random realizations of the normal part Hamiltonian $H$. The results are presented in Fig. \ref{fig:3} (a). The energies are measured relative to the quasiclassical gap energy $E_c$ and are normalized on the corresponding $\Delta_g$. The extent of the smile gap $E_\textrm{sm}$ is thus not visible on the  scale of the Figure, while different values of $\delta_s$ give rise to different Thouless energies and thus to different values of $E_\textrm{sm}$. We thus confirm that in the universal energy units the gap distribution does not depend on $\delta_s$, same as  the quasiclassical value of the gap. We observe however that the averages of the distributions are  shifted  by several $\Delta_g$  from the quasiclassical value of the gap edge.

Next, we fix the level spacing to $\delta_s=0.4\Delta$ and compute the gap distributions for several different values of $N$ ranging from $N=150$ to $N=600$. Each distribution was obtained from $10^4$ random Hamiltonians generated. The results are presented in Fig. \ref{fig:3} (b). As before, the energies are measured from the quasiclassical gap edge $E_c$ and normalized to the particular value of $\Delta_g$. The strong differences in the distributions are due to $N$-dependent shifts.  The shapes of the distributions are indistinguishable with numerical accuracy. This behavior is similar to that of the mesoscopic minigap fluctuations \cite{vavilov:01}. To account for  the shifts, we fit the average values with   $\Delta_\textrm{shift}(N)=a+b N^{2/3}$,  $a$ and $b$ being the fitting parameters. The fit is within several per cent as shown in  Fig. \ref{fig:3} (c). 
The constant term $a$ in this expression reminds  the constant shift  for the universal minigap distributions.  The term $\sim N^{2/3}$ should come from the $N^{2/3}$-scaling of the quasiclassical gap $E_\textrm{sm}$: Not only $E_\textrm{sm}$ in units $\Delta_g$ scales like $N^{2/3}$, but also the average mesoscopic shifts have this scaling. In the limit $N \to \infty$ the first term can be neglected in comparison with the $N^{2/3}$-term

With these shifts, the distributions for different $N$ are in perfect agreement, as shown in  Fig. \ref{fig:3} (d).

\subsection{Gap distribution in the limit $M \gg N \gg 1$}

\begin{figure}[t] 
 \includegraphics[width=0.95\columnwidth,angle=0]{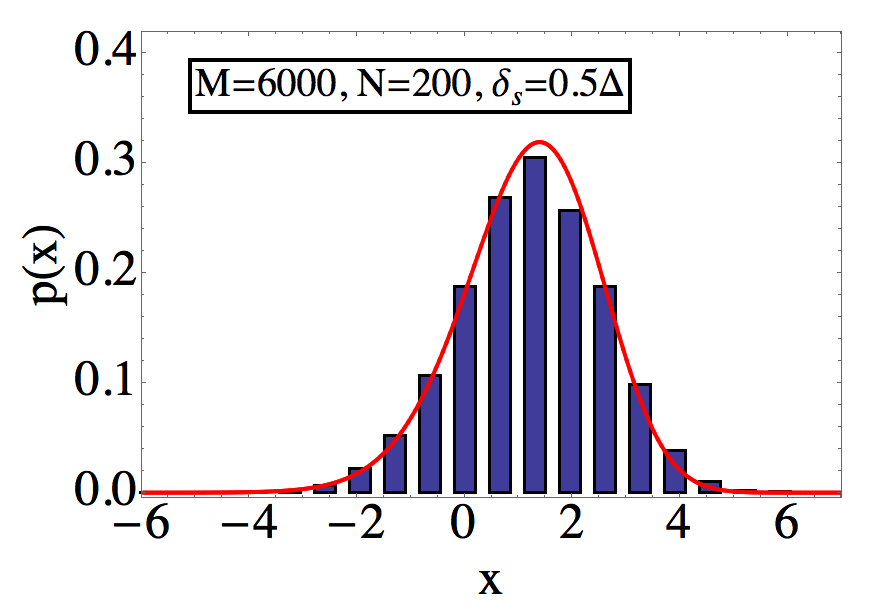}
\caption{\label{fig:4} The distribution of the secondary gap for $M=6000$ and $N=200$ with a level spacing of $\delta_s=0.5\Delta$ compared to the universal distribution found in \cite{vavilov:01} for the minigap in the regime $E_\textrm{Th}\ll \Delta$, where the system can be described by an effective Hamiltonian. The average of the numerical distribution was shifted, as described in the text, in order to reach good agreement with the universal curve. This is possible since the average value seems to be the only cumulant having a strong dependence on $M$ and $N$.}
\end{figure}

It remains unclear if the finite $M/N=5$ distributions presented in the previous subsection are close to the distributions in the limit $M\to \infty$ and $N \to \infty$. In this subsection, we investigate this considering the limit $E_{\textrm{Th}} \gg \Delta$ at bigger ratios $M/N$. We approach the task as follows: For a fixed value of $N$ we calculate the distribution of the gap for a range of large $M$. To combine this with large $E_{\textrm{Th}}$, we fix the level spacing to $\delta_s=0.5\Delta$. We have learned from the previous results that the distributions collapse at this $\delta_s$. The largest values of $M$ and $N$ we are able to reach are $M=6000$ and $N=200$. The resulting distribution of the rescaled gap width $x=(\Delta-E_c)/\Delta_g$
is shown in Fig. \ref{fig:4}. The red curve in Fig. \ref{fig:4} shows the universal distribution function \cite{Tracy:96}, derived in \cite{vavilov:01} for the distribution of the minigap. We observe the agreement between our numerical data and the universal distribution, although the system under consideration cannot be reduced to an effective Hamiltonian. To reach the agreement, we shift the average of our numerical data, $x\to x-x^\star$. This is justified by the results of \cite{vavilov:01}, where the main effect of finite $M$ and $N$ was shown to be a shift of averages not affecting the shape of the distribution.

\begin{figure}[t] 
\begin{overpic}[width=0.98\columnwidth,angle=0]{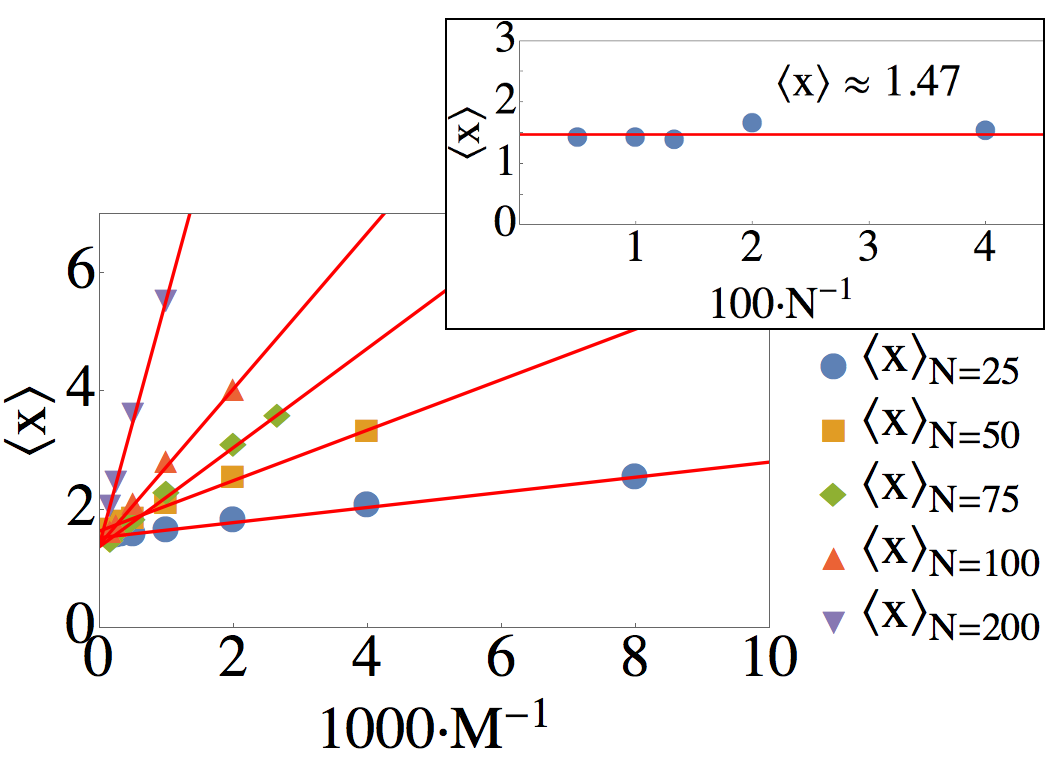}
\put(3,70){\makebox(0,3){$(a)$}}
\end{overpic}
\begin{overpic}[width=0.98\columnwidth,angle=0]{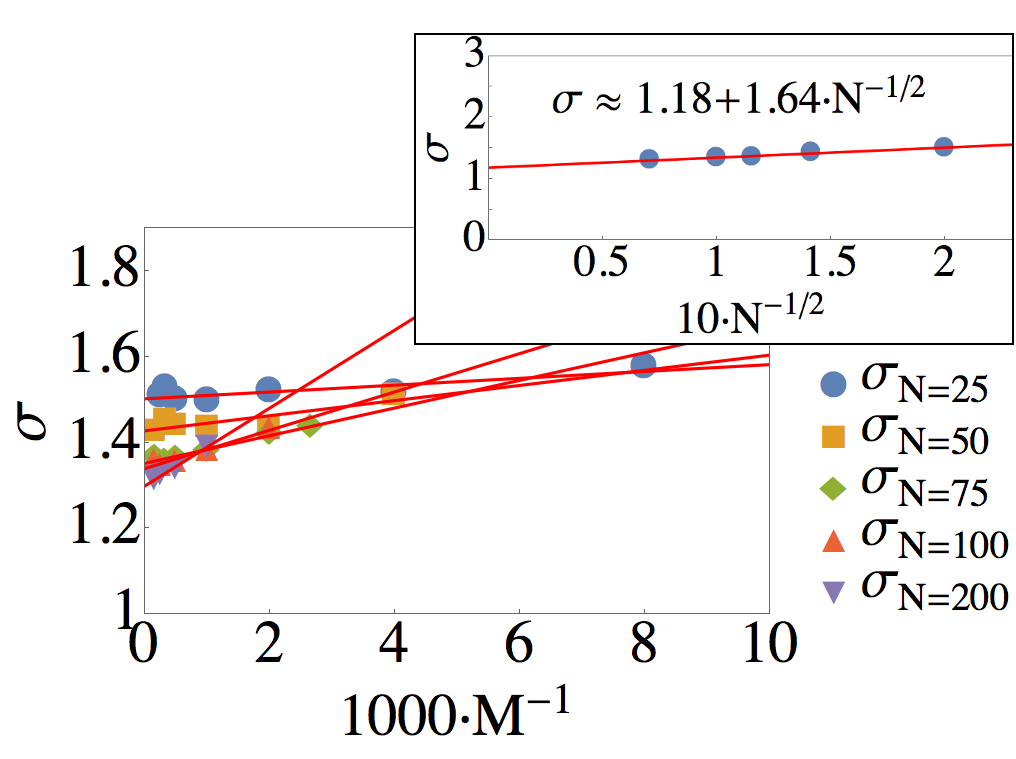}
\put(3,70){\makebox(0,3){$(b)$}}
\end{overpic}
\caption{\label{fig:5} Averages (a) and variances (b) of the secondary gap distribution for different values of $N$ ranging from $N=25$ to $N=200$ as a function of $M^{-1}$. The fit of a straight line for each $N$ allows to estimate the limit $M \to \infty $. Averages as well as variances seem to be only weakly $N$-dependent. Especially for the averages the crossing points with the $y$-axis seem to coincide. The insets in (a) and (b) contain plots of the $M \to \infty $ extrapolated values of $\langle x\rangle$ and $\sigma$ for finite $N$ as a function of $N^{-1}$ and $N^{-1/2}$ respectively. The averages seem to be almost constant as a function of $N$. The $N$-dependence of $\sigma$ is well approximated by fitting a straight line to the data. This allows to estimate the limit $N \to \infty$. The functional form of the fits are given in both cases. For the limit $M\to\infty$ and $N \to \infty$ we estimate the first two cumulants as $\langle x\rangle_{univ.}\approx 1.47$ and $\sigma_{univ}\approx 1.18$, which are quite close to the first two cumulants of the universal curve of Fig. \ref{fig:4}}
\end{figure}

To quantify the agreement even further, we compute the average $\langle x \rangle$ and the variance $\sigma$ for finite $M$ and $N$ and  extrapolate to the limit $M \to \infty$. We repeat this for different values of $N$, and finally extrapolate to the  limit $N \to \infty$. The results are presented in Fig. \ref{fig:5}, where we consider the sets  $N=25$, $N=50$, $N=75$, $N=100$ and $N=200$. For each value of $N$ we calculate the distribution of the gap for different values of $M$, where we always choose $M\gg N$. From these distributions we calculate the averages $\langle x\rangle$ (Fig. \ref{fig:5} (a)) and variances (Fig. \ref{fig:5} (b)) and plot them as a function of $M^{-1}$. Employing the linear fit, we determine the $M \to \infty$ limit from the crossing of the fit with the $y$-axis. At the second step, we fit these results for finite $N$ with a linear fit, assuming $N^{-1}$ and $N^{-1/2}$ corrections for the average and variance, respectively. The fits are shown in insets of Fig. \ref{fig:5}.  From this procedure, we find  $\langle x\rangle_{{\rm univ}}\approx 1.47$ and $\sigma_{{\rm univ}}\approx 1.18$. This we need to compare with the cumulants of the universal distribution: $\sigma_{{\rm univ}}\approx 1.27$ and $\langle x\rangle_{{\rm univ}}\approx 1.21$. We observe the correspondence within $10 \%$ for the variation. The discrepancy in $\langle x\rangle_{{\rm univ}}$ is about two times larger and can be attributed to the uncertainty in the shifts.

\section{Conclusion}
\label{sec:conclusions}
In conclusion, we have studied the statistics of the secondary gap, so-called smile gap, in the spectrum of superconducting nanostructures. We employ a random matrix model. However, the Andreev levels in this case cannot be directly associated with eigenstates of a single random Hamiltonian and are determined from the roots of a spectral determinant. Its construction involves two matrices: an $M\times M$ matrix representing the normal-state region, and $N\times M$ matrix representing its connection to the superconducting leads ($M>N$). While computing the ``smile'' gap width distribution for finite matrices, we have found that the finite matrix dimensions $M$ and $N$ strongly influence only the average of this distribution, while its shape is hardly sensitive to their concrete values; the distribution becomes universal in the limit $ M,N\to\infty $. This way, we have demonstrated that the statistics of the smile gap edge satisfy the universal Tracy-Widom distribution for the edge of an RMT eigenvalue spectrum. 

Remarkably, the statistics of the width of the ``smile'' gap, which we investigated in the limit  $E_\textrm{Th} \gg \Delta$, is the same as the one found in~\cite{vavilov:01} for a more familiar minigap formed at $E_\textrm{Th} \ll \Delta$ around the Fermi level. In distinction from~\cite{vavilov:01}, the universal energy scale $\Delta_g$ for the fluctuations of the ``smile'' gap depends not only on the average level-spacing $\delta_s$: instead, it is a combination of the width of the ``smile'' gap $E_g$ and the level spacing $\delta_s$. 

Our findings emphasize the universality of the Tracy-Widom distribution in the spectrum of superconducting nanostructures with very different origins forming the edge of a quasi-continuous spectrum. 
It will be interesting to establish the connections of our findings with the universal singularities occurring at the gap closures in the random-matrix theory uncovered in~\cite{Brezin:98}, as well as with the phase transitions within the random-matrix theory description of large-$N$ lattice gauge theories~\cite{Gross:80}.



\section*{Acknowledgment}
 J.~R. and W.~B. were supported by the Carl Zeiss Foundation. Yu.~N. acknowledges the support by the European Research Council (ERC) under the European Union's Horizon 2020 research and innovation programme (grant
agreement No. 694272). L.~G. acknowledges the support by NSF DMR-2002275.


\end{document}